\documentclass{vldb}

\usepackage{times}
\usepackage{cite}
\usepackage{graphicx}
\usepackage{amssymb}
\usepackage{mathptm}
\usepackage{setspace}
\usepackage{epsfig}

\newcommand{\eat}[1]{}
\newtheorem{definition}{Definition}

\newtheorem{example}{Example}

\usepackage{lipsum}

\begin{document}

\title{Datom: Towards modular data management}

\numberofauthors{1} 

\author{
%
%
\alignauthor
Verena Kantere
\\
       \affaddr{Institute of Services Science}\\
       \affaddr{Centre Universitaire d' Informatique}\\
        \affaddr{University of Geneva}\\
       \email{verena.kantere@unige.ch}
}

\maketitle
\begin{abstract}
\eat{Data management has been tackling problems in a heterogeneous, isolated, ad hoc and, frequently, solely empirical - based manner. }Recent technology breakthroughs have enabled data collection of unprecedented scale, rate, variety and complexity that has led to an explosion in data management requirements. Existing theories and techniques are not adequate to fulfil these requirements. We endeavour to rethink the way data management research is being conducted\eat{, and work towards a holistic approach of data management problems. 
W} and we propose to work towards modular data management that will allow for unification of the expression of data management problems and systematization of their solution. The core of such an approach is the novel notion of a \emph{datom}, i.e. \emph{a data management atom}, which encapsulates generic data management provision. The datom is the foundation for comparison, customization and re-usage of data management problems and solutions. 
The proposed approach can signal a revolution in data management research and a long anticipated evolution in data management engineering. \eat{Modular data management can leverage data management from its middle ages to its renaissance, by enabling multi-faceted provision, synthesis and re-usage of data management solutions as well as mining of the limitations and determination of the boundaries of data management unification and systematization.}

\end{abstract}




\section{Introduction}

In the last few years, there has been an explosion in data management requirements in various technological environments. The reason is the abundance of electronic devices, the extreme capabilities of scientific instruments and tools, the ubiquitous nature of modern computing, and the notion of offering computing as a service through computing clouds. These breakthroughs have enabled data collection of unprecedented scale, rate, variety and complexity. Even though such collection is already a reality, existing theories and techniques are not adequate to support such data management. 

Trying to cope with the new reality of data management requirements, current research has expanded its efforts to tackle a wider, than the traditional, range of problems. Traditional data management problems have considered a limited range of constraints, assumptions and optimization goals. Typically, the inputs of problems are data requests (i.e. what does the user need to know from the data) and data updates (how do the data change, usually with respect to time); the assumptions are that the computing resources (i.e. CPU, I/O, storage space) and environment (i.e. network communication and workload) are static; and optimization aims to, in general, time-efficiency (i.e. data processing time). Beyond these, modern problems take as inputs computing costs and user budgets, as well as any type of data management guarantees (e.g. privacy, processing deadlines); the assumptions of the new problems include elasticity of computing resources and a dynamic \eat{computing environment }(e.g. number of users and changing data and workload features) and the optimization goals include, beyond time-efficiency, cost, profit, reliability, scalability, etc. 

While there are significant efforts to solve the new types of problems, current techniques have not been able to do so effectively, because they do not yet offer the following: full functionality of traditional DBMSs (e.g., support for transactions, optimization techniques) as transparent services on data; clear description of data management functional parameters (e.g. query execution flow) and guarantees (e.g. cost and risk); near real-time data management; automation; online data processing; transparent management of both data and metadata, homogeneous data manipulation for several data models (e.g. relational, xml, files); migration of data management and adaptation on different infrastructures (e.g. a mobile device, a PC and a cloud provider), etc.  Furthermore, existing research offers one-time solutions, developed for a specific problem in hand and unsuitable to be systematically reused, following the traditional research methodology, which tackles problems in a heterogeneous, isolated, ad hoc and, frequently, solely empirical-based manner.

We argue that the data management world needs to rethink the way it is and has been conducting research, and work towards a holistic approach of modern data management problems. This fundamental research should focus on how data management should be provisioned in order to achieve \emph{optimality}, \emph{unification}, \emph{systematization} and \emph{reusability} of data management solutions. The new era of multi-dimensional data management requirements makes this revision absolutely necessary and urgent. 

\begin{figure}[t!]
\centering\includegraphics [scale = 0.23]{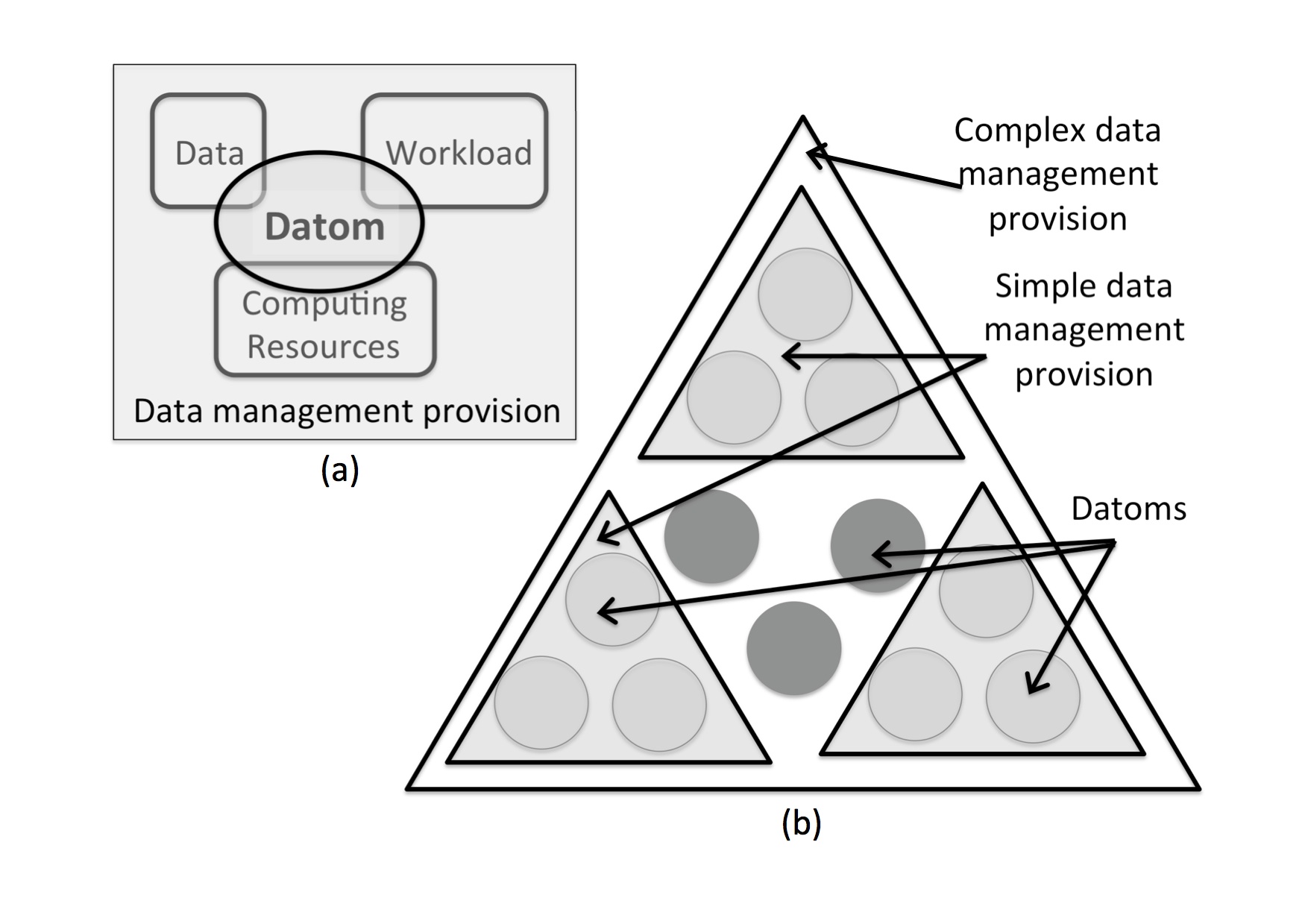}
 \vspace{-0.3in}
   \caption{\label{figure1} (a) Datoms are units of data management provision combining data, workload and computing resources. (b) Data management provision can be synthesized hierarchically on top of datoms and simpler datom syntheses.}
  \vspace{-0.2in}
\end{figure}

Towards this goal, we propose modular data management, which will allow for unification of the expression of data management problems and systematization of their solution. Modular data management can make data management provision the first-class citizen of data management problems, i.e. abstract data management qualities and handle them in a systematic and concrete manner. The core of such an approach is the novel notion of \emph{datom}, i.e. \emph{data management atom}, which encapsulates generic data management provision in terms of data, workload and computing resources, (Figure \ref{figure1}(a)). 
The datom is the foundation for comparison, customization and re-usage of data management problems and solutions. Datoms express self-contained data management provision and can be synthesized hierarchically in more complex entities to express complex data management needs and capabilities (Figure \ref{figure1}(b)). Their creation, customization and synthesis can be automated in order to achieve near-real time data management provision.

The proposed approach can signal a revolution in data management research and a long anticipated evolution in data management engineering, by enabling \eat{multi-objective optimization of data management provisioning} synthesis and re-usage of of data management provisioning solutions as well as mining of the limitations and determination of boundaries of data management systematization. 

In the following: Section \ref{stateoftheart} discusses the state-of-the-art. Section \ref{problem} introduces the approach of modular data management, Section \ref{datom} discusses the notion of datom, Section \ref{discussion} concludes with a discussion.\vspace{-0.05in}

\section{State-of-the-art}\label{stateoftheart}

We discuss the role of modular data management in modern data management and its relation to traditional data management.

\textbf{Cloud data management.} Cloud computing is the ideal paradigm for the management of big amounts of data, which is offered transparently to the user as a service. Cloud data management reshuffles traditional and recent optimization objectives, such as efficiency, fault-tolerance, parallelization, reliability, with new conditions on workload and computing facilities, related to dynamicity, multi-tenancy, elasticity and resource shareness \cite{DBLP:journals/debu/Abadi09}. Also, it adds new optimization objectives with respect to cost and management guarantees. Research has focused mainly on the elasticity and shareness of resources, but also the role of cost \cite{specialissue}. Modular data management aims to complement such efforts by providing methods and tools for the re-usage of one-time solutions in an environment with elastic resources and volatile data and workload. Furthermore, it will enable solutions with multiple and changing optimization objectives, depending on the particularities of the cloud environment. 

\textbf{Big and scientific data management}. The notion of big data is attracting the attention of the data management world, as it is expected that data itself should and will guide decision-making. The persistent data management issues of heterogeneity, scale, timeliness, complexity, and privacy \cite{big data} become insurmountable. Modular data management will contribute to the efficient and automated processing of heterogeneous data, by offering decomposition of complex queries across various data sources and integration of query results.
Scientific data management suffers from lack of automation, online processing, data and process integration \cite{Ailamaki:2010:MSD:1743546.1743568}. Currently, scientists need to collaborate tightly with computer engineers to develop custom solutions that efficiently support data storage and analysis for different experiments \cite{Gray:2005:SDM:1107499.1107503}. Constant collaboration of multidisciplinary scientists and engineers is hard, time and effort consuming, and, furthermore, the gained experience is not disseminated in the scientific community, so that new experimentation can benefit from it. It is necessary to develop generic solutions for storage and analysis of scientific data that can be extended and customized easily. Modular data management  will allow for the extraction of common low-level procedures, customizable for scientific analysis, and synthesizable, in an automated fashion, for scientific workflows.
 
\textbf{Multi-query optimization.} In traditional data management, optimization has been explored with respect to the execution of single and, furthermore, multiple queries. Single-query optimization deals with (time-)efficient execution of one query by building alternative plans using data structures, transforming plans and selecting one for execution \cite{Chaudhuri:1998:OQO:275487.275492, Jarke84queryoptimization}. Multi-query optimization deals with the problem of optimizing the execution of a set of queries in terms of (time-)efficiency. Existing research creates global optimal plans searching exhaustively \cite{Sellis:1988:MO:42201.42203}, or discovers and materializes common query sub-expressions to reduce the cumulative execution cost \cite{Roy:2000:EEA:342009.335419}, does not scale to large numbers of queries and focuses on minimization of execution time. Part of modular data management aims to tackle multi-query optimization in a broader and more generic manner: It will enable the exploration of multi-query optimization on multiple dimensions, related to data, workload and computing characteristics. Also, it will enable the proposal of solutions that are flexible and adaptable to changes of such characteristics, as well as to changes of the optimization goals. Modula data management focuses on the discovery and exploitation of common sub-problems rather than common query logical sub-expressions.
 
\section{Modular data management}\label{problem}

Cloud, big and scientific data management lack of (i) profiling and (ii) provisioning planning, of data management. The overall solution of this problem necessitates a framework for creating and handling data management in a modular and incremental manner.

\textbf{Motivating example.} Situations of data provisioning as the following occur in all three domains:\\
1.	Execute query $Q_A$ on dataset $D_1$ stored and managed by computing resources $C_1$ in a way that is both time-, T, and monetary cost-, C, efficient. To solve this problem, we need to know the alternative execution plans of $Q_A$ on $D_1$/$C_1$, and their efficiency in terms of T and C. Therefore, we  need to define the meaning of Ôtime- and cost-efficiencyÕ: Do we need a pareto-optimal solution $min_x$$[T(x), C(x)]$, where $x$ are the decision variables dependent on replication and data-structure building policies, or do we need to create a weighted aggregate objective function? Also, we need to study the dependency of cost and time.
\\
2.	We need to resolve a situation similar to problem (1). Examples: \\
\indent a.	Execute query $Q_A$ on dataset $D_2$ managed by resources $C_2$, both time-, $T$, and cost-, $C$, efficiently. \\
\indent b.	Execute query $Q_B$ on dataset $D_1$ managed by resources $C_1$, both time-, $T$, and cost-, $C$, efficiently. \\
We need to know if and how we can reuse the solution of problem (1) in order to build the solution of problems (2a) and (2b). This means that we need to know about the relation of alternative query execution plans in the two situations and if and how this relation affects the optimization of time and cost. \\
3.	We need to solve the combination of problems (1) and (2): i.e. either execute $Q_A$ on both settings $D_1$/$C_1$ and $D_2$/$C_2$ or execute both $Q_A$ and $Q_B$ on $D_1$/$C_1$ and integrate the results. So, we need to know if and how we can combine the solutions of problems (1) and (2).\\
4.	We need to execute a query on setting $D_1$/$C_1$, but the optimization objective is more complex or the problem includes constraints, e.g. we want query execution to be both T- and C-efficient while we can guarantee availability of results when 1/3 of the data is processed. In this case we need to know if and how we can build the solution of the problem on the solution of problem (1).\\
The overall solution of the above problems includes the steps:\\
\indent (a)	Formalize the problems so that they can be related to each other and other similar problems.\\
\indent (b)	Create an abstracted and generalized problem; study and solve the generalized problem.\\
\indent (c)	Form the problem inputs, constraints, assumptions, goals and solution as data management objects that can be synthesized and re-used for the solution of similar or more complex problems.

\eat{
\subsection{Problem challenges}

There are two major challenges in unifying and systematizing data management provisioning. The first is a challenge of semantics: It is unclear what exactly is considered as data management provision and to what extent it can be, or it is meaningful, to concretely describe and measure it, and how. This challenge becomes even harder to tackle for data management that incorporates any user requirements that may arise in modern computing environments. The second is a challenge of complexity: There is an explosion in the number of dimensions of modern data management requirements. These dimensions are currently of unknown profile and unknown respective metrics.  

Beyond the above, there are two more challenges. Current computing environments need dynamically provided data management almost at real-time. This dictates simplification of the problem in hand and search for approximate solutions. Yet, experience has shown that data management solutions are solutions of high stakes, meaning that, failing to give an appropriate optimization to a data management problem, may even harm data management beyond the point of operating without any optimization (for example, building and using a data index in order to accelerate query execution may result in even decelerating the later, compared to not using any index, if the built index is not appropriate). Thus, we need to be extremely careful not to oversimplify the problems or over-approximate the given solutions.}


\textbf{Proposed solution.} In this new perspective, data management provision should be the first-class citizen of data management problems. The focus should be on defining what data management is in terms of data, workload and computing resources and enabling the unified and formal definition of data management problems. We propose to achieve these goals through modular data management provision, i.e. \emph{data management provision disseminated into self-contained pieces that can be manipulated atomically and synthesized as groups}. We propose the notion of \emph{datom}, i.e. \emph{data management atom} which refers to a generic unit of data management provision. The datom captures basic data management, in terms of both needs and capabilities, and expresses it in a way that allows for comparison, customization and reusage in order to systematically realize complex data management provision.
Based on the novel notion of datom, we can create methodologies for the realization of the above, depending on the peculiarities of each data management area. Within these methodologies we can create methods and techniques that aim at fulfilling data management requirement.

\section{The datom}\label{datom}\vspace{-0.05in}

 
\subsection{Defining datoms}
 
Feasible data management can be divided in basic entities of data management that can be reused as a whole and in combination with other entities in order to express data management needs and capabilities, and eventually matched in order to cover needs with capabilities. We call such data management units, \emph{datoms}, since they are not dissected further, they are manipulated atomically, and they are employed in combination to create data management solutions.\vspace{-0.05in}

\begin{definition}
A datom $A$ is a set $A = \{P, O, X\}$ where $P$ is a set of properties $P = \{D, W, C\}$. $D$ is the set of properties referring to data, $W$ is the set of properties referring to workload and $C$ is the set of properties referring to computing resources. $O$ is a set of operators $O = \{o_1, \dots, o_m\}$ that take as operands members of $P$, and $X$ is a set of axioms defined on $P$.
\end{definition}\vspace{-0.05in}

A datom has properties concerning all three elements. For example, a datom can represent: the execution of a SPJ (i.e. select-project-join) query with one join, on an attribute with selectivity $10\%$ on a table of 10M tuples running on $1$ CPU with no data transfer via network. Datoms can be supersets/subsets of, or can overlap with, other datoms in terms of properties, operators or axioms. For example, the previous datom is a subset of the datom that corresponds to: the execution of a SPJ query with one join, on an attribute with selectivity $10\%$ on a table of 10M tuples running on $1$ CPU with no data transfer via network using an index on the join attribute.  
 
The complexity and variety of problems that need to be expressed necessitate the creation of a big variety of datoms. Since many of them may be similar, we could create template datoms, or else, semi-constructed datoms, that can be customized for a specific problem. 
Datoms include operators to manipulate property values and axioms to coordinate such a manipulation. Operators and axioms should be inherent to datoms, in order for the latter to be self-contained and employed transparently for complex data management provision.

\subsubsection{Datom properties}\vspace{-0.05in}

The properties of datoms can describe constraints and assumptions, and solutions of problems. Thus, they describe characteristics of data, workload and computing resources. Each of these elements has qualities that are of interest to data management. For example:

\textbf{Data} are characterized by: replication degree, update rate, security constraints, data structures, partitioning degree and type, selectivity estimation, etc.

\textbf{Workload} is characterized by: parallelization of execution, query complexity, data access skewness or similarity, etc.

\textbf{Computing resources} have: CPU utilization, I/O operations, bandwidth, storage space, shareness between real machines, size of virtual machines, etc.

The above characteristics may be dissected to finer ones, e.g. the update rate is either update of existing data or insertion of new information. Furthermore, properties include information about what are the data, workload and computing resources comprised in a datom.

Each property can be measured with one or more metrics. For example, replication degree may be  measured with data size or with tuples/columns to be replicated. Currently, many properties are semantically ambiguous or have different semantics under different data management situations. Thus, there is a necessity for alternative metrics, but also a necessity for the determination of relationships between such metrics.
Moreover, various metrics may have various types, e.g. continuous, discrete, attribute-like. We expect that datom properties will also need  metrics with probabilistic or even possibilistic values, e.g. partitioning may be measured as high and low, update rate may be accompanied by probability values.

Datoms should have properties from the above pool, but it is not necessary for datoms to have all or the same properties. For example, a datom may have the property concerning the data replication degree, but may not have the property concerning the partitioning degree. Such a datom is agnostic to the last property, meaning that it can be customized for any value of the latter.

\subsubsection{Datom operators}\vspace{-0.05in}
Datoms need operators in order to manipulate, i.e. set and expose, the values of their properties. Such operators may be unique to each datom, or common among them. For example, a datom may have an operator to set or exhibit what data or what workload it comprises. Furthermore, it is not mandatory for a datom to have such an operator, meaning that data or workload may be static and hidden. Operators are actually the tools to handle datoms so that we can customize  and combine them. Overall, they enable the reusage  and synthesis of datoms in order to achieve complex modular data management. 

\subsubsection{Datom axioms}\vspace{-0.05in}
The datom operators work under a possible set of axioms that refer to the properties. Specifically, they may refer to the existence of properties, or to their manipulation.
 The axioms of a datom are specific to this datom and, in general, do not hold for other datoms. The role of the axioms is to ensure the atomicity of the datom, meaning that the  datom always represents the data management provision for which it has been designed. \vspace{-0.05in}

\begin{example}
A datom represents a SPJ query with one join, noted as workload $W = \{Q_1\}$,  on some data and computing resources. This datom should never allow for setting another value for $W$, meaning that the axioms should include the following one: $A_1 = \{W \cup Q = W\}$.
\end{example} \vspace{-0.1in}

\subsubsection{Datom qualities}\vspace{-0.05in}
Datoms can be evaluated with respect to the qualities of data management provision they represent. This evaluation, which can be performed analytically or experimentally, depending on the situation, is persistent for each datom and can be employed for reusing the datom in realizing complex data management provision.
Data management qualities can be influenced by characteristics related to:

\textbf{Data integration}, such as data and metadata integration, accuracy with respect to data summarization, annotation, approximation etc. 

\textbf{Performance}, such as efficiency, scalability, availability etc.

\textbf{Privacy}, related to anonymization or security guarantees etc.

\textbf{Cost}, related to execution cost, maintenance cost, money, labor, energy consumption etc.
  
\subsection{Creating datoms}
We can create the datoms either from scratch or reusing existing ones. From-scratch creation necessitates the definition of properties concerning explicit and implicit characteristics. 
For example, a datom that involves the management of the data in a table $R_1$ should include a property $R_1$, but it can also include properties about this data, such as their skewness, $S$ e.g. $S = 10\%$. Creating datoms based on existing ones, can be done only in an additive fashion, meaning that we can only add properties, operators and axioms, as long as the latter do not contradict the existing ones. In such a creation, the original datom works actually as a template for the new ones. This type of datom creation can benefit data management situations that have little change or that reappear in time. \vspace{-0.07in}

\begin{example} \label{examplecreate}
Let us assume there is a datom that refers to querying data in table $R_1$ using some indexes. This datom, $A_a = \{P_a,O_a,X_a\}$, has $D_a =  \{Tables,$ $Indexes\}$ where $Tables = \{R_1\}$ and originally $Indexes = \{ I_1\}$; the datom has an operator that can add new indexes on the data: $O_b = \{O_1(Indexes, I_{new}) = Indexes \cup I_{new}\}$. Furthermore, there can be an axiom that denotes that adding the same index does not make any difference: $A = \{A_1 = I \cup I = I\}$. Let assume that the situation changes and that we need to manage data that reside in two tables $R_1$ and $R_2$ again with the possible association of indexes: This datom, $A_b = \{P_b,O_b,X_b\}$, has $D_b =  D_a = \{Tables, Indexes\}$ where $Tables = \{R_1, R_2\}$ and $Indexes = \{ I_1, I_2, I_{1,2}\}$, where $I_1$ is an index on $R_1$, $I_2$ is an index on $R_2$ and $I_{1,2}$ is an index on both $R_1$ and $R_2$;  $O_b = O_a$ and $A_a = A_b$. \vspace{-0.05in}
\end{example}

\eat{Creating datoms based on other datoms changes the profile of the original datom in terms of data management qualities\eat{, such as traditional ones: efficiency, scalability, availability, and non-traditional ones: security, privacy, integration degree}. }The reason to create datoms based on datoms is for re-usage and comparison purposes: First, if the datom to be created is a superset in terms of properties, operators and/or axioms of existing ones, it is easy and fast to reuse the latter in order to create the first. Second, if a data management situation changes, it may be useful to create the new datoms based on the old ones, in order to be able to compare data management qualities in relationship to the introduced characteristics.
Nevertheless, datoms are meant to be used in compound units, and this we can achieve if we synthesize them to describe complex data management situations. 


\subsection{Synthesizing datoms}
Datoms are meant to describe generic data management provision in a clear and concrete manner. Thereafter, their purpose is to allow for the description of more elaborate data management provision through their synthesis with other datoms. Such synthesis is enabled by datoms through their operators, who allow for customization through setting property values and manipulation \linebreak through the retrieval of property values.

The axioms are local to each datom and guide the synthesis of datoms so that the values and properties of each one stay coherent with the original representation of data management provision by each datom. Nevertheless, since axioms are internal to each datom, it is possible to synthesize datoms with contradicting axioms.

The synthesis of datoms towards the description of complex data management provision necessitates structuring techniques, which are able to build such entities hierarchically, based on single datoms or simpler combinations of datoms. Such techniques should be guided by the overall characteristics, i.e. data, workload and computing resources, but also the overall qualities of the data management provision, i.e. performance or other guarantees, to be achieved. Moreover, these techniques should include tools for comparison of datoms, concerning both data management properties and qualites. The definition of datoms is a first big step towards this direction, since it enables the expression of data management provision in a unified and systematic manner.

Occasionally, datom synthesis may achieve the same result as datom creation ( e.g. through datom customization), but, in general, this is not the case. The reason is that customization invokes changes to the characteristics and/or qualities of data management,which may be not feasible to achieve through synthesis of existing datoms, and therefore, through synthesis of existing characteristics and/or qualities of data management.\vspace{-0.05in}

\begin{example} 
The datom of Example \ref{examplecreate}, $A_a = \{P_a,O_a,$ $X_a\}$ refers to the management of data in $R_1$ using an index $I_1$ and a datom $A_c = \{P_c,O_c,X_c\}$ which refers to the management of data in $R_2$ using an index $I_2$, i.e. $D_c =  \{Tables,$ $Indexes\}$ where $Tables = \{R_2\}$ and $Indexes = \{ I_2\}$. The combination of $A_a$ and $A_c$ is not equivalent with datom $A_b$, since the latter includes also an index $I_{1,2}$. This is rational, as the combined performance of $A_a$ and $A_c$ cannot be equivalent to the performance of $A_b$.\vspace{-0.05in}
\end{example}\vspace{-0.05in}

\section{Discussion}\label{discussion}

\textbf{Possibilities and limitations.} 
Modular data management 
is the foundation for comprehensive solutions to problems with multifarious requirements, constraints and assumptions. Furthermore, it is the foundation for the development of such solutions efficiently in terms of time but also effort, through automation of customization and re-usage of data management modules. This makes feasible the synthesis of ad hoc and near-real-time data management provision, suitable for very dynamic environments such as computing clouds and big data streaming and analytics. 

The challenge is to apply modular data management for problems with ambiguous semantics or peculiar specifics, which may be external to data management. Yet, there can be a big advantage even for such a case: datoms and their manipulation cab abstract and generalize problems, and demonstrate the possible extent of data management unification and systematization.


\textbf{Impact.} 
Modular data management can bridge the gap between research and industry. Providers will be able to design fully functional solutions tailored to each customer. Improved techniques will benefit the processing of business, social and personal data, leading to more effective information extraction and knowledge discovery.
Furthermore, it will provide the foundation for next-generation data management that offers automation, online processing and self-organization. These will achieve the efficient management of tremendously big and fast data collection and processing, which is currently the bottleneck for many science disciplines.
Life experience has shown for more traditional sciences, such as medicine, biology, sociology, psychology etc and other human activities, like music creation, that systematizing knowledge and its discovery, is the answer for boosting advancement in the domain. Likewise, taking this big step in data management, as this will boost the advancement not only in this domain, but to other domains that depend on its progress.

\bibliography{datom_bibliography_shrunk}

\end{document}